\documentclass [11pt,a4paper] {article}

\usepackage[cp1252]{inputenc}

\usepackage{amssymb}
\usepackage{amsmath}
\usepackage{amsfonts,amssymb}
\usepackage[dvips]{graphicx}
\usepackage{bbm}
\usepackage{enumerate}

\usepackage{amsthm}

\usepackage{cancel}

\DeclareMathAlphabet{\mathpzc}{OT1}{pzc}{m}{it}

\def\SmallColSep{\setlength{\arraycolsep}{1pt}}

\begin{document}

\title{Admissibility of truth assignments for quantum propositions in supervaluational logic}

\author{Arkady Bolotin\footnote{$Email: arkadyv@bgu.ac.il$\vspace{5pt}} \\ \textit{Ben-Gurion University of the Negev, Beersheba (Israel)}}

\maketitle

\begin{abstract}\noindent The structure of a complete lattice formed by closed linear subspaces of a Hilbert space (i.e., a Hilbert lattice) entails some unreasonable consequences from the physical point of view. Specifically, this structure seems to contradict to the localized variant of the Kochen-Specker theorem according to which the bivaluation of a proposition represented by a closed linear subspace that does not belong to a Boolean algebra shared by the state, in which a quantum-mechanical system is prepared, must be value indefinite. For this reason, the Hilbert lattice structure seems to be too strong and needs to be weakened. The question is, how should it be weakened so that to support the quantum uncertainty principle and the Kochen-Specker theorem? Which logic will a weakened structure identify? The present paper tries to answer these questions.\\

\end{abstract}

\section{Introduction and preliminaries}  

\noindent The assumption of the structure of a complete lattice imposed on the set $\mathcal{L}(\mathcal{H})$ of closed linear subspaces $\mathcal{H}_a$, $\mathcal{H}_b$, … of a Hilbert space $\mathcal{H}$ (called \textit{Hilbert lattice} \cite{Redei}) entails some unreasonable consequences from the physical point of view.\\

\noindent To see this, let us introduce the proposition $P_a( |\Psi\rangle )$ set forth as follows\smallskip

\begin{equation} \label{DEF} 
   P_a( |\Psi\rangle )
   \equiv
   \mathrm{Prop}
   \left(
      |\Psi\rangle \in \mathcal{H}_{|\Psi\rangle} \wedge \mathcal{H}_a
   \right)
   \;\;\;\;  ,
\end{equation}
\smallskip

\noindent where $\mathrm{Prop}(\cdot)$ is a propositional function (i.e., a sentence that becomes a proposition when all variables in it are given definite values), $\mathcal{H}_{|\Psi\rangle} $  is one of the closed linear subspaces of $\mathcal{L}(\mathcal{H})$ containing the state vector $|\Psi\rangle$ in which a quantum-mechanical system is prepared, $\mathcal{H}_a$ is the subspace of $\mathcal{L}(\mathcal{H})$ that represents the proposition $P_a$, and the subspace $\mathcal{H}_{|\Psi\rangle} \wedge \mathcal{H}_a$ (if it exists) is the meet of $\mathcal{H}_{|\Psi\rangle}$  and $\mathcal{H}_a$ such that\smallskip

\begin{equation}  
   \mathcal{H}_{|\Psi\rangle}, \mathcal{H}_a \in \mathcal{L}(\mathcal{H})
   \;
   \implies
   \;
   \mathcal{H}_{|\Psi\rangle} \wedge \mathcal{H}_a
   =
   \mathcal{H}_{|\Psi\rangle} \!\cap \mathcal{H}_a
   \in \mathcal{L}(\mathcal{H})
   \;\;\;\;  ,
\end{equation}
\smallskip

\noindent where $\cap$ is the set-theoretical intersection. By way of illustration, consider the simplest case of \textit{a qubit}, a two-state quantum-mechanical system such as a one-half spin particle (say, an electron) whose spin may assume only two possible values: either $+\frac{\hbar}{2}$ (denoted as ``up'' or ``$+$'' for short) or $-\frac{\hbar}{2}$ (denoted as ``down'' or ``$-$''). Suppose, the qubit is prepared in the pure state coinciding with the normalized eigenvector $|\Psi^{(z)}_{+}\rangle = [\begin{smallmatrix} 1 \\ 0 \end{smallmatrix}]$ of the Pauli matrix $\sigma^{(z)}$. This eigenvector lies in the range of the projection operator $\hat{P}^{(z)}_{+} = |\Psi^{(z)}_{+}\rangle\langle\Psi^{(z)}_{+}|$, the closed linear subspace of the two-dimensional Hilbert space $\mathcal{H} = \mathbb{C}^2$, namely,\smallskip

\begin{equation}  
   \mathrm{ran}(\hat{P}^{(z)}_{+}\!)
   =
   \left\{
      \!\left[
         \begingroup\SmallColSep
         \begin{array}{r}
            a \\
            0
         \end{array}
         \endgroup
      \right]\!
      :
      \,
      a \in \mathbb{R}
   \right\}
   \;\;\;\;  .
\end{equation}
\smallskip

\noindent According to the definition (\ref{DEF}), in the state $|\Psi^{(z)}_{+}\rangle$ the proposition ``Spin along the $z$-axis is up'', denoted as $P^{(z)}_{+}$ and represented by the range $\mathrm{ran}(\hat{P}^{(z)}_{+})$, is given by\smallskip

\begin{equation}  
   P^{(z)}_{+}( |\Psi^{(z)}_{+}\rangle )
   =
   \mathrm{Prop}
   \!
   \left(
      |\Psi^{(z)}_{+}\rangle
      \in
      \mathrm{ran}(\hat{P}^{(z)}_{+}\!)
      \wedge
      \mathrm{ran}(\hat{P}^{(z)}_{+}\!)
   \right)
   =
   \mathrm{Prop}
   \!
   \left(
      |\Psi^{(z)}_{+}\rangle
      \in
      \mathrm{ran}(\hat{P}^{(z)}_{+}\!)
   \right)
   \;\;\;\;  ,
\end{equation}
\smallskip

\noindent which is \textit{a tautology}. By contrast, in the same state, the proposition ``Spin along the $z$-axis is down'', denoted as $P^{(z)}_{-}$ and represented by the range\smallskip

\begin{equation}  
   \mathrm{ran}(\hat{P}^{(z)}_{-}\!)
   =
   \left\{
      \!\left[
         \begingroup\SmallColSep
         \begin{array}{r}
            0 \\
            a
         \end{array}
         \endgroup
      \right]\!
      :
      \,
      a \in \mathbb{R}
   \right\}
   \;\;\;\;   
\end{equation}
\smallskip

\noindent containing the second normalized eigenvector $|\Psi^{(z)}_{-}\rangle = [\begin{smallmatrix} 0 \\ 1 \end{smallmatrix}]$  of the Pauli matrix $\sigma^{(z)}$, is given by\smallskip

\begin{equation}  
   P^{(z)}_{-}( |\Psi^{(z)}_{+}\rangle )
   =
   \mathrm{Prop}
   \!
   \left(
      |\Psi^{(z)}_{+}\rangle
      \in
      \left\{
         \!\left[
            \begingroup\SmallColSep
            \begin{array}{r}
               a \\
               0
            \end{array}
            \endgroup
         \right]\!
         :
         a \in \mathbb{R}
      \right\}
      \cap 
      \left\{
         \!\left[
            \begingroup\SmallColSep
            \begin{array}{r}
               0 \\
               a
            \end{array}
            \endgroup
         \right]\!
         :
         a \in \mathbb{R}
      \right\}
   \right)
   \!
   =
   \mathrm{Prop}
   \!
   \left(
      |\Psi^{(z)}_{+}\rangle
      \in
      \{ 0 \}      
   \right)
   \;\;\;\;  ,
\end{equation}
\smallskip

\noindent which is \textit{a contradiction} due to $|\Psi\rangle \neq 0$, i.e., the statement that any physically meaningful state of the quantum system differs from vector $0$. In a bivalent semantics (defined by \textit{the bivaluation relation}, i.e., the function $b$ from the set of propositions into the set $\{0,1\}$ of bivalent truth values) this can be expressed equivalently, using $0$ for \textit{false} and $1$ for \textit{true}, in the following way:\smallskip

\begin{equation}  
   b
   \left(
      P^{(z)}_{+}( |\Psi^{(z)}_{+}\rangle )
   \right)
   =
   b
   \left(
      \mathrm{Prop}
      \!
      \left(
         |\Psi^{(z)}_{+}\rangle
         \in
         \mathrm{ran}(\hat{P}^{(z)}_{+}\!)
      \right)
   \right)
   \!
   =
   1
   \;\;\;\;  ,
\end{equation}

\begin{equation}  
   b
   \left(
      P^{(z)}_{-}( |\Psi^{(z)}_{+}\rangle )
   \right)
   =
   b
   \left(
      \mathrm{Prop}
      \!
      \left(
         |\Psi^{(z)}_{+}\rangle
         \in
         \{ 0 \}
      \right)
   \right)
   \!
   =
   0
   \;\;\;\;  .
\end{equation}
\smallskip

\noindent Now, recall that in the Hilbert lattice structure, every pair of the elements $\mathcal{H}_a$ and $\mathcal{H}_b$ from $\mathcal{L}(\mathcal{H})$ has the meet \cite{Davey}. As a result, in the state $|\Psi^{(z)}_{+}\rangle$, the propositions ``Spin along the $x$-axis is up'' and ``Spin along the $x$-axis is down'' (denoted as $P^{(x)}_{+}$ and $P^{(x)}_{-}$, correspondingly, and represented by the ranges $\mathrm{ran}(\hat{P}^{(x)}_{+})$ and $\mathrm{ran}(\hat{P}^{(x)}_{-})$ containing in that order the normalized eigenvectors $|\Psi^{(x)}_{+}\rangle = \frac{1}{\sqrt{2}} [\begin{smallmatrix} 1 \\ 1 \end{smallmatrix}]$  and $|\Psi^{(x)}_{-}\rangle = \frac{1}{\sqrt{2}} [\begin{smallmatrix} \;\;\:1 \\ -1 \end{smallmatrix}]$  of the Pauli matrix $\sigma^{(x)}$) are \textit{determined}\smallskip

\begin{equation}  
   \mathrm{ran}(\hat{P}^{(z)}_{+}\!)
   \wedge
   \mathrm{ran}(\hat{P}^{(x)}_{+}\!)
   =
   \left\{
      \!\left[
         \begingroup\SmallColSep
         \begin{array}{r}
            a \\
            0
         \end{array}
         \endgroup
      \right]\!
      :
      a \in \mathbb{R}
   \right\}
   \cap 
   \left\{
      \!\left[
         \begingroup\SmallColSep
         \begin{array}{r}
            a \\
            a
         \end{array}
         \endgroup
      \right]\!
      :
      a \in \mathbb{R}
      \right\}
   =
   \{ 0 \}
   \;\;\;\;  ,
\end{equation}

\begin{equation}  
   \mathrm{ran}(\hat{P}^{(z)}_{+}\!)
   \wedge
   \mathrm{ran}(\hat{P}^{(x)}_{-}\!)
   =
   \left\{
      \!\left[
         \begingroup\SmallColSep
         \begin{array}{r}
            a \\
            0
         \end{array}
         \endgroup
      \right]\!
      :
      a \in \mathbb{R}
   \right\}
   \cap 
   \left\{
      \!\left[
         \begingroup\SmallColSep
         \begin{array}{r}
             a \\
            -a
         \end{array}
         \endgroup
      \right]\!
      :
      a \in \mathbb{R}
   \right\}
   =
   \{ 0 \}
   \;\;\;\;  ,
\end{equation}
\smallskip

\noindent and so they have a bivalent truth value:\smallskip

\begin{equation} \label{X12} 
   b
   \left(
      P^{(x)}_{\pm}( |\Psi^{(z)}_{+}\rangle )
   \right)
   =
   b
   \left(
      \mathrm{Prop}
      \!
      \left(
         |\Psi^{(z)}_{+}\rangle
         \in
         \mathrm{ran}(\hat{P}^{(z)}_{+}\!)
         \wedge
         \mathrm{ran}(\hat{P}^{(x)}_{\pm}\!)         
      \right)
   \right)
   \!
   =
   0
   \;\;\;\;  .
\end{equation}
\smallskip

\noindent However, by the postulates of quantum mechanics, measurements of spin along the different axes are incompatible. Since this requires that both propositions $P^{(x)}_{+}$ and $P^{(x)}_{-}$ be \textit{undetermined} in the state where the proposition $P^{(z)}_{+}$ is verified (i.e., has the value of the truth), quantum logic (i.e., the complete orthomodular lattice based on the closed subspaces of a Hilbert space \cite{Mackey, Ptak}) seems to contradict to the quantum mechanical uncertainty principle.\\

\noindent Furthermore, quantum logic seems to contradict to \textit{the localized variant of the Kochen-Specker theorem} \cite{Abbott} stating that the bivaluation of the proposition $P_a(|\Psi\rangle)$, represented by the closed linear subspace $\mathcal{H}_a$ that does not belong to a Boolean algebra shared by the state $|\Psi\rangle$ in which the quantum-mechanical system is prepared, must be value indefinite (e.g., instead of (\ref{X12}) it must be $b( P^{(x)}_{\pm}( |\Psi^{(z)}_{+}\rangle ) ) = \frac{0}{0}$, where $\frac{0}{0}$ denotes an indeterminate value).\\

\noindent Therefore, the Hilbert lattice structure seems to be too strong and thus needs to be weakened. The question is, how should it be weakened so that to support the uncertainty principle and the Kochen-Specker theorem? Which logic does a weakened structure identify?\\

\noindent The present paper tries to answer these questions.\\

\section{From supervaluationary logic to invariant-subspace lattices}  

\noindent It was already observed in \cite{Chiara} that the Hilbert lattice structure could be weakened by the requirement for the meet operation to exist only for the subspaces belonging to a common Boolean algebra. However, it was not clear what semantics would associate with such a requirement. Let us show that it is \textit{supervaluationism}.\\

\noindent Recall that supervaluationism is a semantics in which a proposition can have a bivalent truth value even when its components do not. Therefore, it allows one to apply the tautologies of propositional logic in cases where bivalent truth values cannot be defined \cite{Varzi, Keefe}.\\

\noindent Take, for example, \textit{the Heisenberg cut}, i.e., the borderline that seemingly separates the quantum realm from the classical \cite{Zeh, Schlosshauer, Duvenhage}. This cut ensures that sufficiently small systems will manifest superpositions and hence quantum behavior; on the other hand, for sufficiently large systems, superpositional (quantum) behavior will be replaced by classical behavior. It leads to the question of where exactly it should be, where a classical system begins and a quantum system ends. It relates to another question of when the collapse of the wave function takes place and how long it takes (since this is essentially the same Heisenberg cut but with space replaced by time) \cite{Hajicek}.\\

\noindent Since the concept of a classical system appears to lack sharp boundaries and because of the subsequent indeterminacy surrounding the extension of the predicate ``\textit{is a classical system}'', no quantum particle (such as electron, protons, or neutron) constituting a composite system can be identified as making the difference between \textit{being a classical system} and \textit{not being a classical system}. Given then that one quantum particle does not make a classical system, it would seem to follow that two do not, thus three do not, and so on. In the end, no number of the quantum particles can constitute a classical system. This is a paradox since apparently true premises through seemingly uncontroversial reasoning lead to an apparently false conclusion.\\

\noindent To resolve this paradox (known in the literature as \textit{the Sorites Paradox} \cite{Keefe, Hyde}), one may deny that there are such things as classical systems, i.e., the world is quantum rather than classical.\\

\noindent A different approach to this paradox is offered by supervaluationism. Since it is logically true for any number $N$ of the quantum particles that it either does or does not make a classical system, the disjunction of the propositions $P$ = ``$N$ quantum particle(s) constitute(s) a classical system'' and $\neg P$ = ``$N$ quantum particle(s) do(es) not constitute a classical system'' is an instance of the valid schema. That is, $P \vee \neg P$ should be true regardless of whether its disjuncts can be described as either true or false. Otherwise stated, while it is true that there is some borderline separating everything governed by the wave function from a classical description, there is no particular number of the quantum particles $N$ for which it is true that it is the borderline (i.e., the Heisenberg cut). This implies that a semantics for a logic of propositions relating to quantum systems should not be truth-functional (that is, some propositions in this semantics may have no bivalent truth values which is called \textit{truth-value gaps} \cite{Beziau}).\\

\noindent As supervaluationism admits truth-value gaps, to interpret propositions relating to the quantum system in terms of a supervaluationary logic, \textit{a lattice structure allowing truth-value gaps} should be imposed on the closed linear subspaces of the Hilbert space associated with the quantum system. A natural candidate for such a structure is the collection of invariant-subspace lattices that have no mutual nontrivial members.\\

\noindent Recall that a subspace $\mathcal{H}_a \subseteq \mathcal{H}$ is called \textit{an invariant subspace under the projection operator $\hat{P}_a$ on a finite Hilbert space $\mathcal{H}$} if\smallskip

\begin{equation}  
   \hat{P}_a\!
   :\,
   \mathcal{H}_a
   \mapsto
   \mathcal{H}_a
   \;\;\;\;  .
\end{equation}
\smallskip

\noindent Concretely, the image of every vector $|\Psi\rangle$ in $\mathcal{H}_a$ under $\hat{P}_a$ remains within $\mathcal{H}_a$ which can be denoted as\smallskip

\begin{equation}  
   \hat{P}_a \mathcal{H}_a
   =
   \left\{
      |\Psi\rangle \in \mathcal{H}_a
      :\;
      \hat{P}_a |\Psi\rangle
   \right\}
   \subseteq
   \mathcal{H}_a
   \;\;\;\;  .
\end{equation}
\smallskip

\noindent For example, the range and kernel of the projection operator $\hat{P}_a$, i.e., $\mathrm{ran}(\hat{P}_a)$ and $\mathrm{ker}(\hat{P}_a) = \mathrm{ran}(\hat{1} - \hat{P}_a)$, as well as the trivial subspaces $\mathrm{ran}(\hat{0}) = \{0\}$ and $\mathrm{ran}(\hat{1}) = \mathcal{H}$ (where $\hat{0}$ and $\hat{1}$ stand for the trivial projection operators) are the invariant subspaces under $\hat{P}_a$.\\

\noindent Let $\mathcal{L}(\hat{P}_a)$ denote the set of the subspaces invariant under $\hat{P}_a$, namely,\smallskip

\begin{equation}  
   \mathcal{L}(\hat{P}_a)
   \equiv
   \left\{
      \mathcal{H}_a \subseteq \mathcal{H}
      :\;
      \hat{P}_a \mathcal{H}_a
      \subseteq
      \mathcal{H}_a
   \right\}
   \;\;\;\;  .
\end{equation}
\smallskip

\noindent Recall that the family $\Sigma^{(Q)}$ of two or more nontrivial projection operators $\hat{P}^{(Q)}_n$, namely,\smallskip

\begin{equation}  
   \Sigma^{(Q)}
   \equiv
   \left\{
      \hat{P}^{(Q)}_n
   \right\}
   \;\;\;\;  ,
\end{equation}
\smallskip

\noindent is called \textit{a context} if the next two conditions hold:\smallskip

\begin{equation}  
   \hat{P}^{(Q)}_n, \hat{P}^{(Q)}_{m \neq n} \in \Sigma^{(Q)}
   \;
   \implies
   \;
   \hat{P}^{(Q)}_n \hat{P}^{(Q)}_{m \neq n}
   =
   \hat{P}^{(Q)}_{m \neq n} \hat{P}^{(Q)}_n
   =
   \hat{0}
   \;\;\;\;  ,
\end{equation}

\begin{equation}  
   \sum_{\hat{P}^{(Q)}_n \in \Sigma^{(Q)}} \hat{P}^{(Q)}_n
   =
   \hat{1}
   \;\;\;\;  .
\end{equation}
\smallskip

\noindent Consider the set of the invariant subspaces \textit{invariant under each $\hat{P}^{(Q)}_n \in \Sigma^{(Q)}$}:\\

\begin{equation}  
   \mathcal{L}(\Sigma^{(Q)})
   \equiv
   \!\!\!
   \bigcap_{\hat{P}^{(Q)}_n \in \Sigma^{(Q)}}
   \!\!\!
      \mathcal{L}(\hat{P}^{(Q)}_n)
   \;\;\;\;  .
\end{equation}
\smallskip

\noindent For the qubit where $\hat{P}^{(Q)}_{+} \hat{P}^{(Q)}_{-} = \hat{P}^{(Q)}_{-} \hat{P}^{(Q)}_{+} = \hat{0}$ and $\hat{P}^{(Q)}_{+} + \hat{P}^{(Q)}_{-} = \hat{1}$, this set is\smallskip

\begin{equation}  
   \mathcal{L}(\Sigma^{(Q)})
   =
   \mathcal{L}(\hat{P}^{(Q)}_{+})
   \cap
   \mathcal{L}(\hat{P}^{(Q)}_{-})
   =
   \left\{
      \mathrm{ran}(\hat{0})
      ,\,
      \mathrm{ran}(\hat{P}^{(Q)}_{+})
      ,\,
      \mathrm{ran}(\hat{P}^{(Q)}_{-})
      ,\,
      \mathrm{ran}(\hat{1})
   \right\}
   \;\;\;\;  .
\end{equation}
\smallskip

\noindent The elements of every set $\mathcal{L}(\Sigma^{(Q)})$ form a complete lattice called \textit{the invariant-subspace lattice of the context $\Sigma^{(Q)}$} \cite{Radjavi}. It is straightforward to see that each invariant-subspace lattice only contains the subspaces belonging to the mutually commutable projection operators, that is, each $\mathcal{L}(\Sigma^{(Q)})$ is a Boolean algebra.\\

\noindent Recall that two contexts are called \textit{intertwined} if they share one or more common elements \cite{Svozil}. As any intertwined context has at least one individual element (i.e., one that is not shared by other contexts), each lattice $\mathcal{L}(\Sigma^{(Q)})$ has a nonempty set of individual subspaces. Clearly, in the structure of the invariant-subspace lattices $\mathcal{L}(\Sigma^{(Q)})$, $\mathcal{L}(\Sigma^{(R)})$, … imposed on the closed linear subspaces of the Hilbert space, the individual subspaces of the different lattices cannot meet each other. In symbols,\smallskip

\begin{equation}  
   \mathrm{ran}(\hat{P}_n^{(Q)})
   \in
   \mathcal{L}(\Sigma^{(Q)})
   ,\,
   \mathrm{ran}(\hat{P}_m^{(R)})
   \in
   \mathcal{L}(\Sigma^{(R)})
   \,
   \implies
   \,
   \mathrm{ran}(\hat{P}_n^{(Q)})
   \;\cancel{\;\wedge\;}\;
   \mathrm{ran}(\hat{P}_m^{(R)})
   \;\;\;\;  ,
\end{equation}
\smallskip

\noindent where $\mathrm{ran}(\hat{P}_n^{(Q)})$ and $\mathrm{ran}(\hat{P}_m^{(R)})$ denote the individual subspaces from the different lattices (i.e., $Q \neq R$), and the cancelation of the meet operation $\wedge$ indicates that this operation cannot be defined for such subspaces (recall that the meet is defined as an operation on pairs of elements from the same partially ordered set $\mathcal{L}$ \cite{Davey}).\\

\noindent The nonexistence of the meet operation for pairs of the ranges that do not lie in a common invariant-subspace lattice corresponds to truth-value gaps in supervaluational logic. To be sure, let the quantum-mechanical system be prepared in the state $|\Psi^{(Q)}_n\rangle$ residing in $\mathrm{ran}(\hat{P}_n^{(Q)})$, the individual subspace from the lattice $\mathcal{L}(\Sigma^{(Q)})$. In this case, the proposition $P^{(R)}_m$ represented by the individual subspace $\mathrm{ran}(\hat{P}_m^{(R)})$ from the lattice $\mathcal{L}(\Sigma^{(R)})$ is undetermined and thus cannot have a bivalent truth value:\smallskip

\begin{equation}  
   b
   \left(
      P^{(R)}_{m}( |\Psi^{(Q)}_{n}\rangle )
   \right)
   =
   b
   \left(
      \mathrm{Prop}
      \!
      \left(
         |\Psi^{(Q)}_{n}\rangle
         \in
         \mathrm{ran}(\hat{P}_n^{(Q)})
         \;\cancel{\;\wedge\;}\;
         \mathrm{ran}(\hat{P}_m^{(R)})
      \right)
   \right)
   \!
   =
   \frac{0}{0}
   \;\;\;\;  .
\end{equation}
\smallskip

\noindent For example, in the state where the proposition ``Spin along the $z$-axis is up'' can be described as either true or false, both propositions ``Spin along the $x$-axis is up'' and ``Spin along the $x$-axis is down'' are neither true nor false: $b( P^{(x)}_{\pm}( |\Psi^{(z)}_{\pm}\rangle ) ) = b( \mathrm{Prop}( |\Psi^{(z)}_{\pm}\rangle \in \mathrm{ran}(\hat{P}^{(z)}_{+}) \;\cancel{\;\wedge\;}\; \mathrm{ran}(\hat{P}^{(x)}_{\pm}) ) ) = \frac{0}{0}$.\\

\noindent Then again, in any invariant-subspace lattice on a finite-dimensional Hilbert space one has\smallskip

\begin{equation}  
   \mathrm{ran}(\hat{P}^{(R)}_m)
   \wedge
   \mathrm{ker}(\hat{P}^{(R)}_m)
   =
   \mathrm{ran}(\hat{0})
   \;\;\;\;  ,
\end{equation}

\begin{equation}  
   \mathrm{ran}(\hat{P}^{(R)}_m)
   \vee
   \mathrm{ker}(\hat{P}^{(R)}_m)
   =
   \left( \mathrm{ran}(\hat{0}) \right)^{\perp}
   =
   \mathrm{ran}(\hat{1})
   \;\;\;\;  ,
\end{equation}
\smallskip

\noindent where $(\cdot)^{\perp}$ stands for the orthogonal complement of $(\cdot)$. Given that the subspaces $\mathrm{ran}(\hat{P}^{(R)}_m)$ and $\mathrm{ker}(\hat{P}^{(R)}_m)$ represent the proposition $P^{(R)}_m$ and its negation $\neg P^{(R)}_m$, while the subspaces $\mathrm{ran}(\hat{0})$ and $\mathrm{ran}(\hat{1})$ represent the conjunction and disjunction of $P^{(R)}_m$ and $\neg P^{(R)}_m$, respectively, one finds\smallskip

\begin{equation}  
   P^{(R)}_m
   \wedge
   \neg P^{(R)}_m
   (|\Psi\rangle)
   =
   \mathrm{Prop}
   \!
   \left(
      |\Psi\rangle
      \in
      \mathcal{H}_{|\Psi\rangle}
      \wedge
      \{0\}
   \right)
   \;\;\;\;  ,
\end{equation}

\begin{equation}  
   P^{(R)}_m
   \vee
   \neg P^{(R)}_m
   (|\Psi\rangle)
   =
   \mathrm{Prop}
   \!
   \left(
      |\Psi\rangle
      \in
      \mathcal{H}_{|\Psi\rangle}
      \wedge
      \mathcal{H}
   \right)
   \;\;\;\;  .
\end{equation}
\smallskip

\noindent Since $\mathrm{Prop}(|\Psi\rangle\!\in\!\mathcal{H}_{|\Psi\rangle} \wedge \{0\})$ is a contradiction and $\mathrm{Prop}(|\Psi\rangle\!\in\!\mathcal{H}_{|\Psi\rangle} \wedge \mathcal{H})$ is a tautology, the conjunction and disjunction of $P^{(R)}_m$ and $\neg P^{(R)}_m$ are always false and true, correspondingly,\smallskip

\begin{equation}  
   b\left(
      P^{(R)}_m
      \wedge
      \neg P^{(R)}_m
      (|\Psi\rangle)
   \right)
   =
   0
   \;\;\;\;  ,
\end{equation}

\begin{equation}  
   b\left(
      P^{(R)}_m
      \vee
      \neg P^{(R)}_m
      (|\Psi\rangle)
   \right)
   =
   1
   \;\;\;\;  ,
\end{equation}
\smallskip

\noindent even with undetermined $P^{(R)}_m(|\Psi\rangle)$ and $\neg P^{(R)}_m(|\Psi\rangle)$.\\

\section{Diagrams of lattice structures}  

\noindent To portray different lattice structures of the closed linear subspaces $\mathcal{H}_a$, $\mathcal{H}_b$, … of a Hilbert space, a modified version of a Hasse diagram can be used.\\

\begin{figure}[ht!]
   \centering
   \includegraphics[scale=0.55]{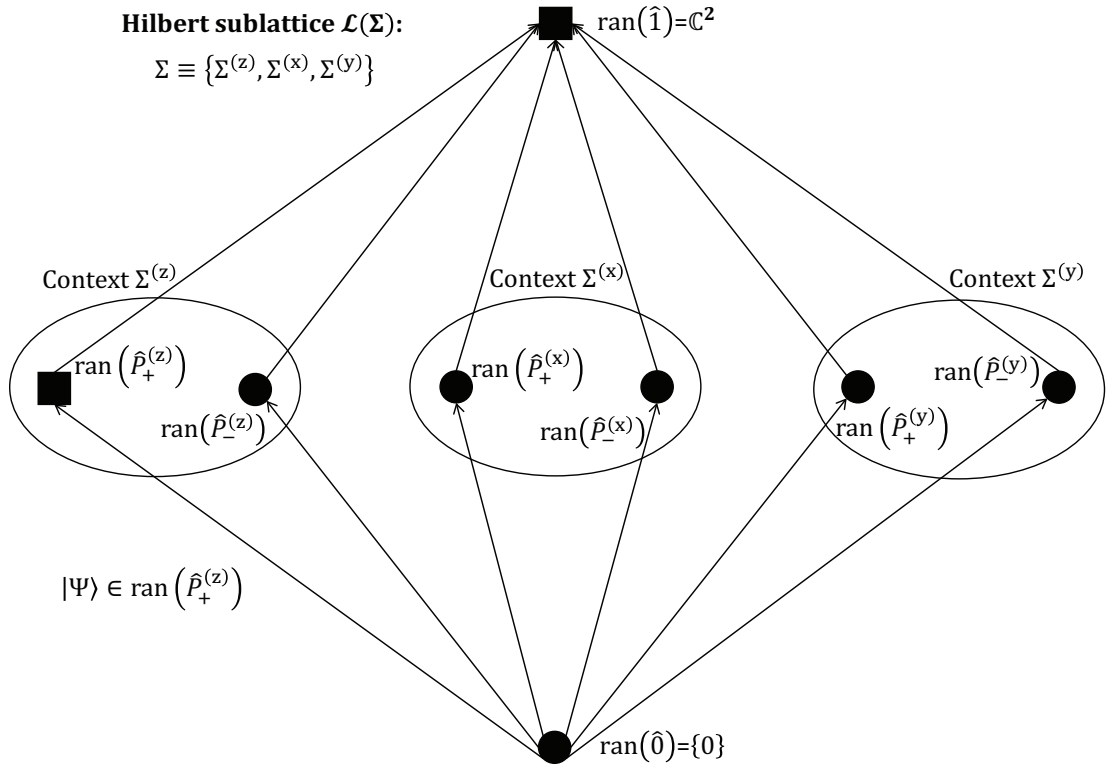}
   \caption{The bivaluation relation in the Hilbert sublattice $\mathcal{L}(\Sigma)$ of the qubit\label{fig1}}
\end{figure}

\noindent Recall that the Hasse diagram is a type of mathematical diagram where each subspace corresponds to a vertex in the plane connected with another vertex by a line segment which goes upward from $\mathcal{H}_a$ to $\mathcal{H}_b$ whenever $\mathcal{H}_a \subseteq \mathcal{H}_b$ and there is no $\mathcal{H}_c$ such that $\mathcal{H}_a \subseteq \mathcal{H}_c \subseteq \mathcal{H}_b$ \cite{Kiena}. Besides the information on the transitive reduction, the Hasse diagram can also show the truth values of the propositions relating to the quantum system in a specific state by picturing the vertices that represent these propositions in the following way: the vertex is drawn as \textit{a black square} if the proposition is \textit{true}, the vertex is drawn as \textit{a black circle} if the proposition is \textit{false}, and the vertex is drawn as \textit{a hollow circle} if the proposition \textit{cannot be described as either true or false}. Moreover, the ellipse-like curves enclosing the subspaces belonging to the common contexts are added to the standard Hasse diagram.\\

\begin{figure}[ht!]
   \centering
   \includegraphics[scale=0.55]{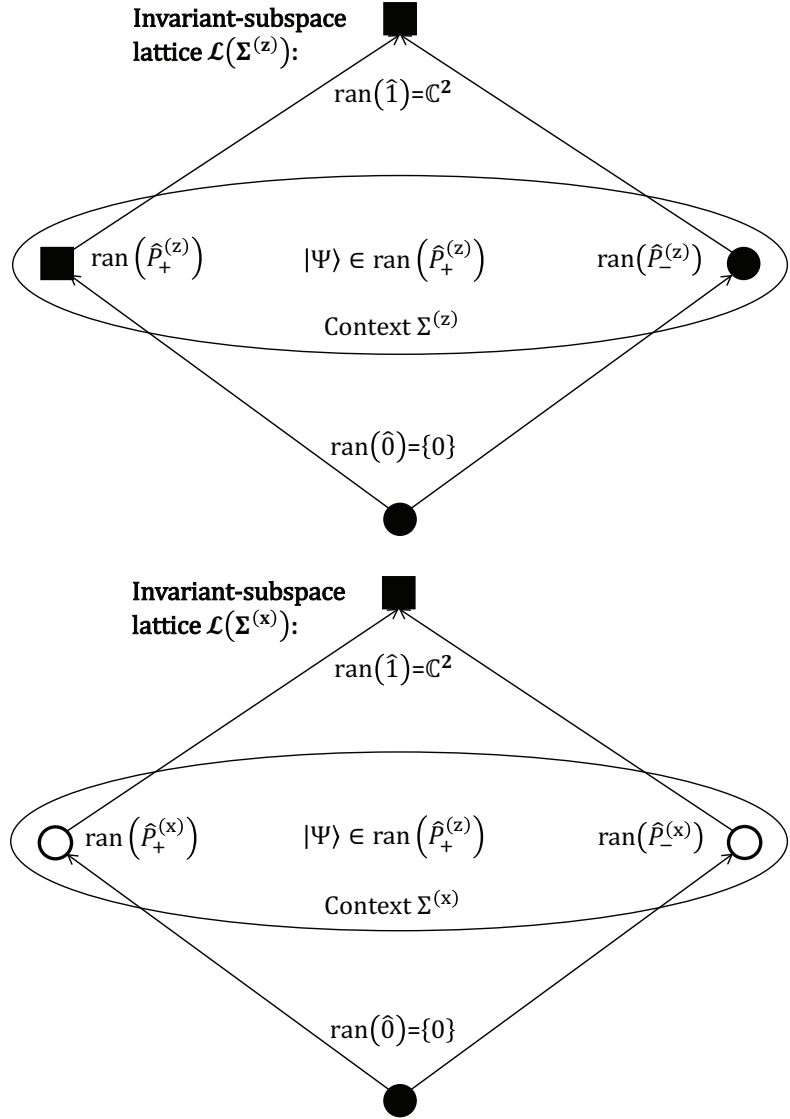}
   \caption{The bivaluation relation in the invariant-subspace lattices of the qubit\label{fig2}}
\end{figure}

\noindent The Figures \ref{fig1} and \ref{fig2} present the diagrams of the structures formed by the closed subspaces of the qubit. Specifically, the Figure \ref{fig1} shows the Hilbert sublattice $\mathcal{L}(\Sigma)$ which is a lattice with the same meet and join operations as the Hilbert lattice $\mathcal{L}(\mathbb{C}^2)$ (the nontrivial closed subspaces of the sublattice $\mathcal{L}(\Sigma)$ correspond to the operators to measure the qubit spin along the $x$, $y$ and $z$ axes), while the Figure \ref{fig2} demonstrates the invariant-subspace lattices $\mathcal{L}(\Sigma^{(z)})$ and $\mathcal{L}(\Sigma^{(x)})$. In both Figures, the truth values are given in the pure state $|\Psi\rangle \in \mathrm{ran}(\hat{P}^{(z)}_{+})$.\\

\noindent Consider three contexts $\Sigma^{(1)}$, $\Sigma^{(2)}$ and $\Sigma^{(6)}$ on the Hilbert space $\mathbb{C}^4$ used in the paper \cite{Cabello} by Cabello et al. to prove the Bell-Kochen-Specker theorem. The projection operators composing these contexts are\smallskip

\begin{equation}  
   \hat{P}^{(1)}_1
   =
   \!\left[
      \begingroup\SmallColSep
      \begin{array}{r r r r}
         0 & 0 & 0 & 0 \\
         0 & 0 & 0 & 0 \\
         0 & 0 & 0 & 0 \\
         0 & 0 & 0 & 1
      \end{array}
      \endgroup
   \right]
   \,
   ,
   \,\,
   \hat{P}^{(1)}_2
   =
   \!\left[
      \begingroup\SmallColSep
      \begin{array}{r r r r}
         0 & 0 & 0 & 0 \\
         0 & 0 & 0 & 0 \\
         0 & 0 & 1 & 0 \\
         0 & 0 & 0 & 0
      \end{array}
      \endgroup
   \right]
   \,
   ,
   \,\,
   \hat{P}^{(1)}_3
   =
   \!
   \frac{1}{2}
   \!\left[
      \begingroup\SmallColSep
      \begin{array}{r r r r}
         1 & 1 & 0 & 0 \\
         1 & 1 & 0 & 0 \\
         0 & 0 & 0 & 0 \\
         0 & 0 & 0 & 0
      \end{array}
      \endgroup
   \right]
   \,
   ,
   \,\,
   \hat{P}^{(1)}_4
   =
   \!
   \frac{1}{2}
   \!\left[
      \begingroup\SmallColSep
      \begin{array}{r r r r}
                  1 & \bar{1} & 0 & 0 \\
         \bar{1} & 1          & 0 & 0 \\
                  0 & 0          & 0 & 0 \\
                  0 & 0          & 0 & 0
      \end{array}
      \endgroup
   \right]
   \;\;\;\;  ,
\end{equation}

\begin{equation}  
   \hat{P}^{(2)}_1
   =
   \!\left[
      \begingroup\SmallColSep
      \begin{array}{r r r r}
         0 & 0 & 0 & 0 \\
         0 & 0 & 0 & 0 \\
         0 & 0 & 0 & 0 \\
         0 & 0 & 0 & 1
      \end{array}
      \endgroup
   \right]
   \,
   ,
   \,\,
   \hat{P}^{(2)}_2
   =
   \!\left[
      \begingroup\SmallColSep
      \begin{array}{r r r r}
         0 & 0 & 0 & 0 \\
         0 & 1 & 0 & 0 \\
         0 & 0 & 0 & 0 \\
         0 & 0 & 0 & 0
      \end{array}
      \endgroup
   \right]
   \,
   ,
   \,\,
   \hat{P}^{(2)}_3
   =
   \!
   \frac{1}{2}
   \!\left[
      \begingroup\SmallColSep
      \begin{array}{r r r r}
         1 & 0 & 1 & 0 \\
         0 & 0 & 0 & 0 \\
         1 & 0 & 1 & 0 \\
         0 & 0 & 0 & 0
      \end{array}
      \endgroup
   \right]
   \,
   ,
   \,\,
   \hat{P}^{(2)}_4
   =
   \!
   \frac{1}{2}
   \!\left[
      \begingroup\SmallColSep
      \begin{array}{r r r r}
                  1 & 0 & \bar{1} & 0 \\
                  0 & 0 & 0          & 0 \\
         \bar{1} & 0 & 1          & 0 \\
                  0 & 0 & 0          & 0
      \end{array}
      \endgroup
   \right]
   \;\;\;\;  ,
\end{equation}
\vspace*{-9mm}

\begin{equation}  
   \hat{P}^{(6)}_1
   =
   \!
   \frac{1}{4}
   \!\left[
      \begingroup\SmallColSep
      \begin{array}{r r r r}
                  1 & \bar{1} & \bar{1} &          1 \\
         \bar{1} &          1 &          1 & \bar{1} \\
         \bar{1} &          1 &          1 & \bar{1} \\
                  1 & \bar{1} & \bar{1} &          1
      \end{array}
      \endgroup
   \right]
   \,
   ,
   \,\,
   \hat{P}^{(6)}_2
   =
   \!
   \frac{1}{4}
   \!\left[
      \begingroup\SmallColSep
      \begin{array}{r r r r}
         1 & 1 & 1 & 1 \\
         1 & 1 & 1 & 1 \\
         1 & 1 & 1 & 1 \\
         1 & 1 & 1 & 1
      \end{array}
      \endgroup
   \right]
   \,
   ,
   \,\,
   \hat{P}^{(6)}_3
   =
   \!
   \frac{1}{2}
   \!\left[
      \begingroup\SmallColSep
      \begin{array}{r r r r}
                  1 &          0 &          0 & \bar{1} \\
                  0 &          0 &          0 &          0 \\
                  0 &          0 &          0 &          0 \\
         \bar{1} &          0 &          0 &          1
      \end{array}
      \endgroup
   \right]
   \,
   ,
   \,\,
   \hat{P}^{(6)}_4
   =
   \!
   \frac{1}{2}
   \!\left[
      \begingroup\SmallColSep
      \begin{array}{r r r r}
                  0 &            0 &            0 &          0 \\
                  0 &            1 &   \bar{1} &          0 \\
                  0 &   \bar{1} &            1 &          0 \\
                  0 &            0 &            0 &          0
      \end{array}
      \endgroup
   \right]
   \;\;\;\;  ,
\end{equation}
\smallskip

\noindent where $\bar{1}$ stands for $-1$.\\

\begin{figure}[ht!]
   \centering
   \includegraphics[scale=0.55]{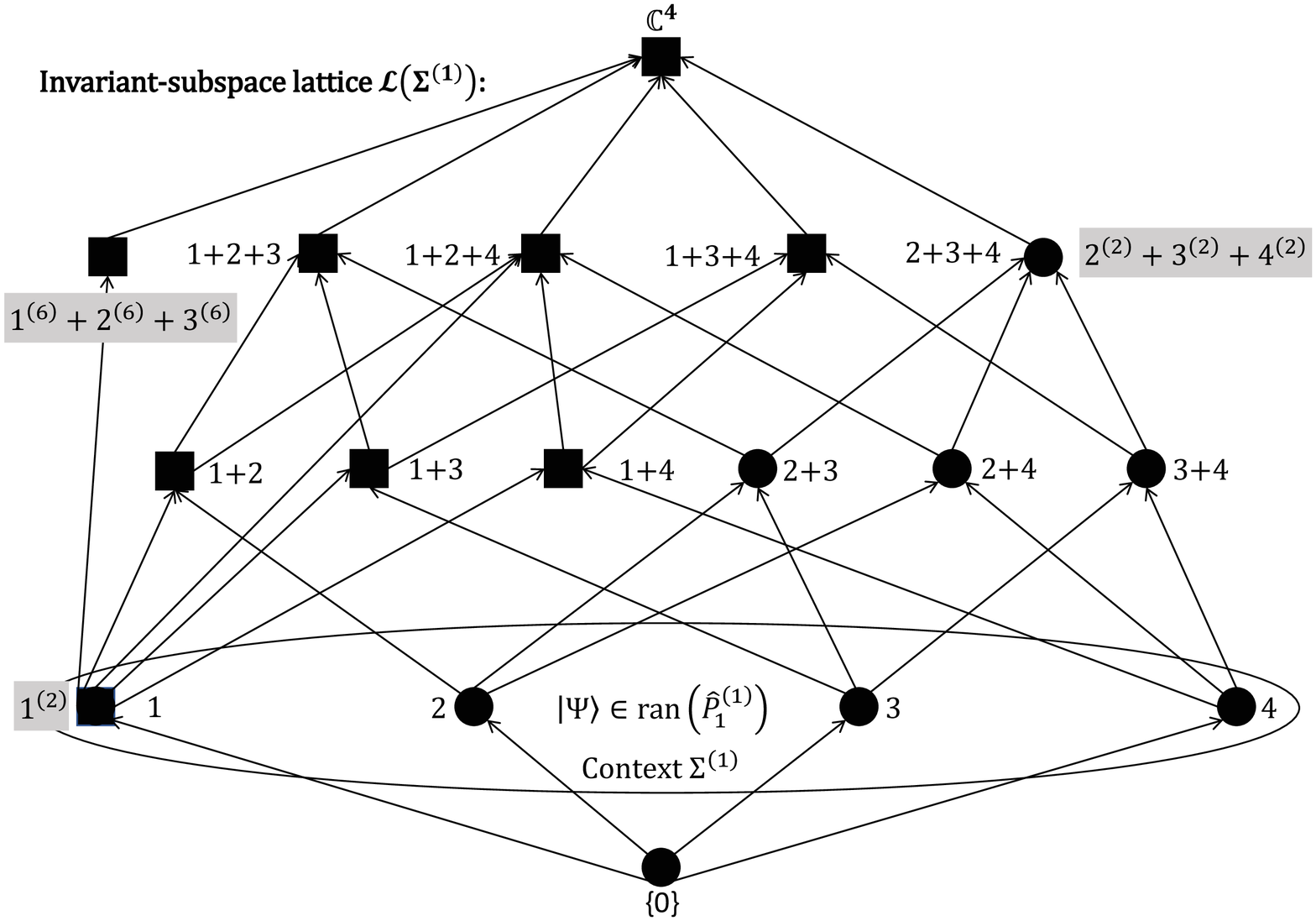}
   \caption{The bivaluation relation in the invariant-subspace lattice $\mathcal{L}(\Sigma^{(1)})$ of Cabello’s set\label{fig3}}
\end{figure}

\begin{figure}[ht!]
   \centering
   \includegraphics[scale=0.55]{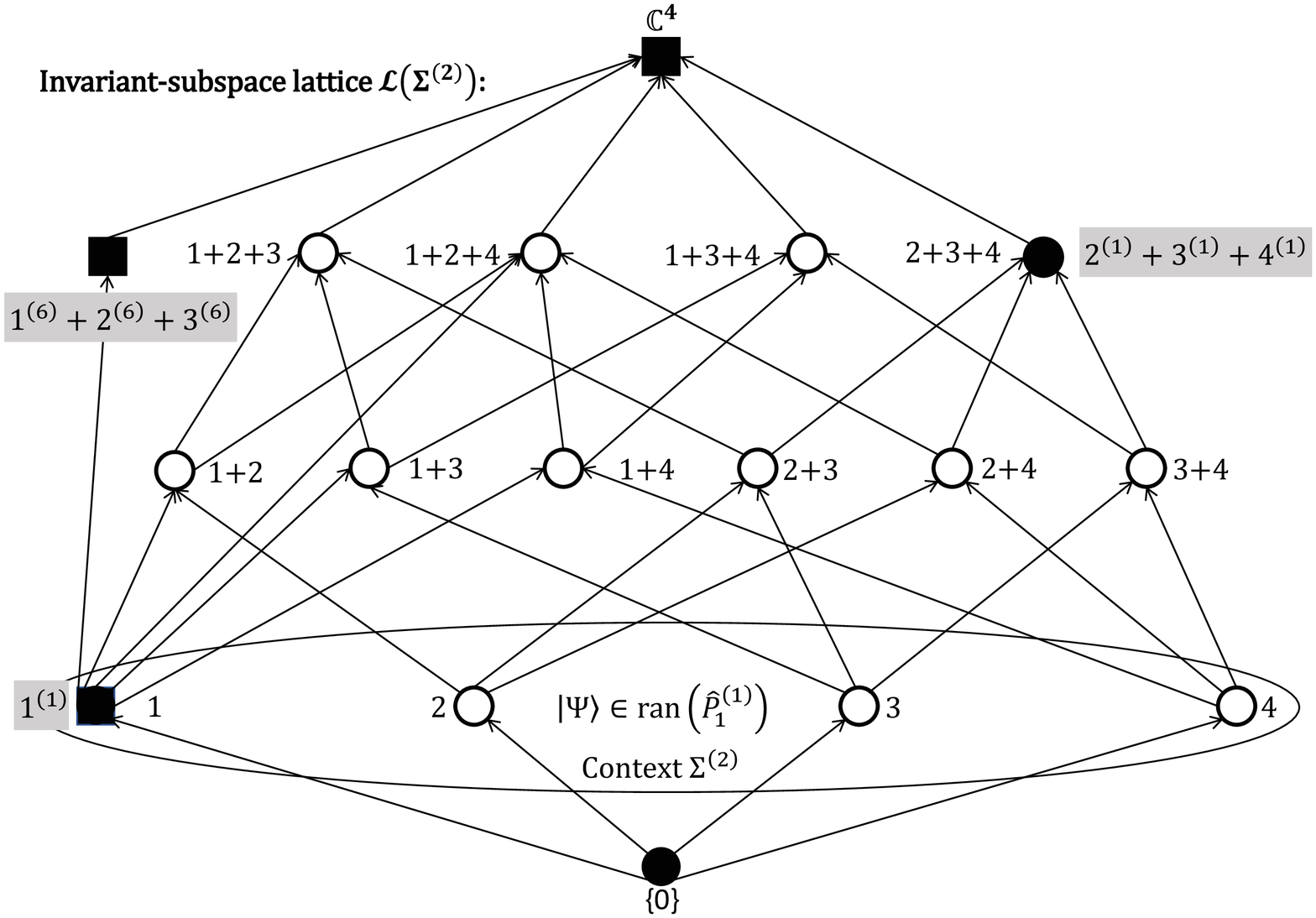}
   \caption{The bivaluation relation in the invariant-subspace lattice $\mathcal{L}(\Sigma^{(2)})$ of Cabello’s set\label{fig4}}
\end{figure}

\begin{figure}[ht!]
   \centering
   \includegraphics[scale=0.55]{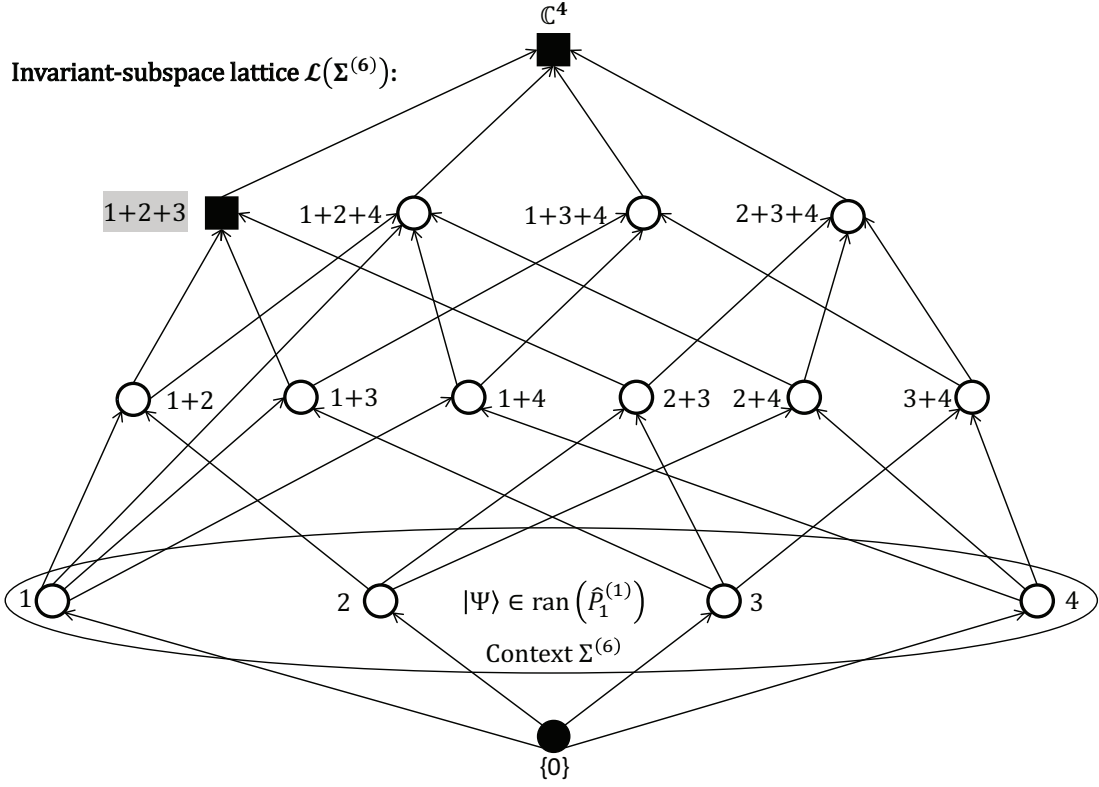}
   \caption{The bivaluation relation in the invariant-subspace lattice $\mathcal{L}(\Sigma^{(6)})$ of Cabello’s set\label{fig5}}
\end{figure}

\noindent The Figures \ref{fig3}, \ref{fig4} and \ref{fig5} show the Hasse diagrams of the invariant-subspace lattices based on the closed subspaces generated by those projection operators in addition to the truth values of the relating propositions given in the pure state $|\Psi\rangle \in \mathrm{ran}(\hat{P}^{(1)}_1)$. In the Figures \ref{fig3}-\ref{fig5}, the following notations are used: ``1'' denotes $\mathrm{ran}(\hat{P}^{(1)}_1)$ (or $\mathrm{ran}(\hat{P}^{(2)}_1)$ or $\mathrm{ran}(\hat{P}^{(6)}_1)$, depending on the lattice), ``1+2'' denotes $\mathrm{ran}(\hat{P}^{(1)}_1 + \hat{P}^{(1)}_2)$, ``1+2+3'' denotes $\mathrm{ran}(\neg \hat{P}^{(1)}_4) = \mathrm{ran}(\hat{P}^{(1)}_1 + \hat{P}^{(1)}_2 + \hat{P}^{(1)}_3)$, and so on.\\

\noindent It is obvious that $\mathrm{ran}(\hat{P}^{(2)}_1)$ and $\mathrm{ker}(\hat{P}^{(2)}_1) = \mathrm{ran}(\neg\hat{P}^{(2)}_1) = \mathrm{ran}(\hat{P}^{(2)}_2 + \hat{P}^{(2)}_3 + \hat{P}^{(2)}_4)$ are the shared closed subspaces belonging to both invariant-subspace lattices $\mathcal{L}(\Sigma^{(2)})$ and $\mathcal{L}(\Sigma^{(1)})$.\\

\noindent What is more, the kernel of the projection operator $\hat{P}^{(6)}_4$ is the invariant subspace under any projection operator from the context $\Sigma^{(1)}$. To be sure,\smallskip

\begin{equation}  
   |\Psi\rangle
   =
   \!
   \!\left[
   \!
      \begingroup\SmallColSep
      \begin{array}{r}
         0 \\
         0 \\
         0 \\
         1
      \end{array}
      \endgroup
      \!
   \right]
   \in
   \mathrm{ker}(\hat{P}^{(6)}_4)
   =
   \!
   \!\left\{
   \!
   \left[
   \!
      \begingroup\SmallColSep
      \begin{array}{r}
         c \\
         b \\
         b \\
         a
      \end{array}
      \endgroup
      \!
   \right]
   \!
   \right\}
   \,
   \implies 
   \,
   \left\{  
   \begin{array}{r}
      \hat{P}^{(1)}_1 |\Psi\rangle
      =
      \!\left[
      \!
         \begingroup\SmallColSep
         \begin{array}{r}
            0 \\
            0 \\
            0 \\
            1
         \end{array}
         \endgroup
         \!
      \right]
      \in
      \!
      \left\{
      \!
      \left[
      \!
         \begingroup\SmallColSep
         \begin{array}{r}
            c \\
            b \\
            b \\
            a
         \end{array}
         \endgroup
         \!
      \right]
      \!
      \right\}
      \\ 
      \hat{P}^{(1)}_{m \neq 1} |\Psi\rangle
      =
      \!\left[
      \!
         \begingroup\SmallColSep
         \begin{array}{r}
            0 \\
            0 \\
            0 \\
            0
         \end{array}
         \endgroup
         \!
      \right]
      \in
      \!
      \left\{
      \!
      \left[
      \!
         \begingroup\SmallColSep
         \begin{array}{r}
            c \\
            b \\
            b \\
            a
         \end{array}
         \endgroup
         \!
      \right]
      \!
      \right\}
   \end{array}
   \right. 
   \;\;\;\;  ,
\end{equation}
\smallskip

\noindent where $a,b,c \in \mathbb{R}$. Hence, $\mathrm{ker}(\hat{P}^{(6)}_4) = \mathrm{ran}(\hat{P}^{(6)}_1 + \hat{P}^{(6)}_2 + \hat{P}^{(6)}_3)$ belongs to both invariant-subspace lattices $\mathcal{L}(\Sigma^{(6)})$ and $\mathcal{L}(\Sigma^{(1)})$. The ordering of $\mathrm{ker}(\hat{P}^{(6)}_4)$ in $\mathcal{L}(\Sigma^{(1)})$ is as follows:\smallskip 

\begin{equation}  
   \mathrm{ran}(\hat{P}^{(1)}_1)
   \cap
   \mathrm{ker}(\hat{P}^{(6)}_4)
   =
   \!
   \!\left\{
   \!
   \left[
   \!
      \begingroup\SmallColSep
      \begin{array}{r}
         0 \\
         0 \\
         0 \\
         a
      \end{array}
      \endgroup
      \!
   \right]
   \!
   \right\}
   \cap 
   \!\left\{
   \!
   \left[
   \!
      \begingroup\SmallColSep
      \begin{array}{r}
         c \\
         b \\
         b \\
         a
      \end{array}
      \endgroup
      \!
   \right]
   \!
   \right\}
   \!
   =
   \!
   \!\left\{
   \!
   \left[
   \!
      \begingroup\SmallColSep
      \begin{array}{r}
         0 \\
         0 \\
         0 \\
         a
      \end{array}
      \endgroup
      \!
   \right]
   \!
   \right\}
   \,
   \implies 
   \,
   \mathrm{ran}(\hat{P}^{(1)}_1)
   \subseteq
   \mathrm{ker}(\hat{P}^{(6)}_4)
   \;\;\;\;  .
\end{equation}
\smallskip

\noindent It can be also shown that $\mathrm{ker}(\hat{P}^{(6)}_4) \in \mathcal{L}(\Sigma^{(2)})$ with the similar ordering.\\

\noindent The text boxes of the shared nontrivial subspaces are colored grey in the diagrams.\\

\section{Conclusion remarks}  

\noindent As it can be seen from the above diagrams, all the propositions represented by the subspaces that do not belong to the invariant-subspace lattices allocated by the state, in which the quantum system is prepared, are undetermined.\\

\noindent This implies that the admissibility rules do not hold true in the structure of the invariant-subspace lattices imposed on the closed linear subspaces in the Hilbert space.\\

\noindent Recall that according to the admissibility rules \cite{Abbott, Svozil}, (1) if among all the subspaces belonging to a context there is one that represents a true proposition, then the others must represent false propositions; what is more, (2) if there is a subspace belonging to a context such that it represents a false proposition, then another subspace in this context must represent either false or true proposition (as long as only one proposition is true).\\

\noindent One can easily check with the presented above diagrams that the second admissibility rule holds true within the structure of the Hilbert sublattice $\mathcal{L}(\Sigma)$ but it does not apply to the structure of the invariant-subspace lattices.\\

\noindent Accordingly, the statement of the Kochen-Specker theorem (expressing that there is no way to assign a bivalent truth value to every quantum proposition relating to a given context such that the admissibility rules hold true \cite{Kochen, Peres}) need not to be proved in supervaluational logic since it is immediately evident in that logic.\\

\bibliographystyle{References}
\bibliography{Admissibility}

\begin{thebibliography}{10}
\expandafter\ifx\csname urlstyle\endcsname\relax
  \providecommand{\doi}[1]{doi:\discretionary{}{}{}#1}\else
  \providecommand{\doi}{doi:\discretionary{}{}{}\begingroup
  \urlstyle{rm}\Url}\fi

\bibitem{Redei}
M.~R$\acute{\mathrm{e}}$dei.
\newblock \emph{Quantum {L}ogic in {A}lgebraic {A}pproach}.
\newblock Springer, Dordrecht, Netherlands, 1998.

\bibitem{Davey}
B.~Davey and H.~Priestley.
\newblock \emph{Introduction to {L}attices and {O}rder}.
\newblock Cambridge University, Cambridge, UK, 2002.

\bibitem{Mackey}
G.~Mackey.
\newblock Quantum {M}echanics and {H}ilbert {S}pace.
\newblock \emph{American Math. Monthly}, 64:45–--57, 1957.

\bibitem{Ptak}
P.~Pt$\acute{\mathrm{a}}$k and S.~Pulmannov$\acute{\mathrm{a}}$.
\newblock \emph{Orthomodular {S}tructures as {Q}uantum {L}ogics}.
\newblock Kluwer, Dordrecht, 1991.

\bibitem{Abbott}
A.~Abbott, C.~Calude, and K.~Svozil.
\newblock A variant of the {K}ochen-{S}pecker theorem localizing value
  indefiniteness.
\newblock \emph{J. Math. Phys.}, 56(102201):1--17, 2015.

\bibitem{Chiara}
M.~Chiara and R.~Giuntini.
\newblock Quantum logic.
\newblock arXiv:quant-ph/0101028, Jan 2001.

\bibitem{Varzi}
A.~Varzi.
\newblock Supervaluationism and {I}ts {L}ogics.
\newblock \emph{Mind}, 116:633--676, 2007.

\bibitem{Keefe}
R.~Keefe.
\newblock \emph{Theories of {V}agueness}.
\newblock Cambridge University Press, Cambridge, 2000.

\bibitem{Zeh}
H.~Zeh.
\newblock Decoherence: {B}asic {C}oncepts and {T}heir {I}nterpretation.
\newblock arXiv:quant-ph/9506020, Jun 2002.

\bibitem{Schlosshauer}
M.~Schlosshauer.
\newblock Decoherence, the measurement problem, and interpretations of quantum
  mechanics.
\newblock \emph{Rev. Mod. Phys.}, 76:1267--1305, 2004.

\bibitem{Duvenhage}
R.~Duvenhage.
\newblock The {N}ature of {I}nformation in {Q}uantum {M}echanics.
\newblock arXiv:quant-ph/0203070, Jun 2006.

\bibitem{Hajicek}
P.~Hajicek.
\newblock Realism-{C}ompleteness-{U}niversality interpretation of quantum
  mechanics.
\newblock arXiv:1509.05547, Sep 2016.

\bibitem{Hyde}
D.~Hyde.
\newblock The sorites paradox.
\newblock In G.~Ronzitti, editor, \emph{Vagueness: a guide}, pages 1--18.
  Springer, 2011.

\bibitem{Beziau}
J.-Y. B{$\grave{\mathrm{e}}$}ziau.
\newblock Bivalence, {E}xcluded {M}iddle and {N}on {C}ontradiction.
\newblock In L.~Behounek, editor, \emph{The {L}ogica {Y}earbook 2003}, pages
  73--84. Academy of Sciences, Prague, 2003.

\bibitem{Radjavi}
H.~Radjavi and P.~Rosenthal.
\newblock \emph{Invariant Subspaces}.
\newblock Dover Publications, 2003.

\bibitem{Svozil}
K.~Svozil.
\newblock Classical versus quantum probabilities and correlations.
\newblock arXiv:1707.08915, Aug 2017.

\bibitem{Kiena}
S.~Kiena.
\newblock Hasse {D}iagrams.
\newblock In \emph{Implementing {D}iscrete {M}athematics: {C}ombinatorics and
  {G}raph {T}heory with {M}athematica}, pages 169--170. Addison-Wesley, 1990.

\bibitem{Cabello}
A.~Cabello, J.~Estebaranz, and G.~Garcia-{A}lcaine.
\newblock Bell-{K}ochen-{S}pecker theorem: {A} proof with 18 vectors.
\newblock \emph{Phys. Letters A}, 212:183--187, 1996.

\bibitem{Kochen}
S.~Kochen and E.~Specker.
\newblock The problem of hidden variables in quantum mechanics.
\newblock \emph{J. Math. Mech.}, 17(1):59--87, 1967.

\bibitem{Peres}
A.~Peres.
\newblock Two simple proofs of the {K}ochen-{S}pecker theorem.
\newblock \emph{Phys. A: Math. Gen.}, 24:175--178, 1991.

\end{thebibliography}

\end{document}